\documentclass[aps,superscriptaddress,twocolumn,twoside,floatfix,prl,nofootinbib,a4paper]{revtex4-2}
\pdfoutput=1
\usepackage{times}
\usepackage{dsfont}
\usepackage{bm} 
\usepackage{amsfonts}
\usepackage{amsmath}
\usepackage{amssymb}
\usepackage{amsthm}
\usepackage{color}
\usepackage{enumerate}
\usepackage{multirow}
\newcommand{\stkout}[1]{\ifmmode\text{\sout{\ensuremath{#1}}}\else\sout{#1}\fi}
\usepackage{latexsym}
\usepackage{mathrsfs}
\usepackage{verbatim}
\usepackage{float}
\usepackage{pifont}
\usepackage{caption}
\usepackage{empheq}
\usepackage{IEEEtrantools}
\usepackage{microtype}
\usepackage[dvipsnames]{xcolor}
\usepackage{natbib}
\usepackage[colorlinks=true,linkcolor=blue,citecolor=magenta,urlcolor=blue]{hyperref}

\usepackage{cleveref} 
\usepackage[many]{tcolorbox}
\newtcolorbox[auto counter]{framefloat}[2][]{title=Box~\thetcbcounter: #2,,fonttitle=\bfseries, boxsep=0mm,boxrule=1pt,colframe=black,colback=white,coltitle=black,float=t!,#1}
\definecolor{maroon}{cmyk}{0,0.87,0.68,0.32}
\DeclareMathOperator{\Tr}{tr}

\usepackage{physics}

\newcommand{\expect}[1]{\langle#1\rangle}

\newcommand{\degree}{{}^{\circ}}

\usepackage{array,mathtools,amssymb,booktabs}
\newcolumntype{C}{>{$}c<{$}}
\AtBeginDocument{
\heavyrulewidth=.08em
\lightrulewidth=.05em
\cmidrulewidth=.03em
\belowrulesep=.65ex
\belowbottomsep=0pt
\aboverulesep=.4ex
\abovetopsep=0pt
\cmidrulesep=\doublerulesep
\cmidrulekern=.5em
\defaultaddspace=.5em
}
\usepackage{epsfig}
\usepackage{graphicx}
\graphicspath{{figures/}}
\usepackage{subfig}
\usepackage{caption}
\begin{document}

\title{Recycling nonlocality in a quantum network}

\author{ Ya-Li Mao}
\altaffiliation{These authors contributed equally to this work.}
\affiliation{Shenzhen Institute for Quantum Science and Engineering and Department of Physics, Southern University of Science and Technology, Shenzhen, 518055, China}
\affiliation{Guangdong Provincial Key Laboratory of Quantum Science and Engineering, Southern University of Science and Technology, Shenzhen, 518055, China}

\author{Zheng-Da Li}
\altaffiliation{These authors contributed equally to this work.}
\affiliation{Shenzhen Institute for Quantum Science and Engineering and Department of Physics, Southern University of Science and Technology, Shenzhen, 518055, China}
\affiliation{Guangdong Provincial Key Laboratory of Quantum Science and Engineering, Southern University of Science and Technology, Shenzhen, 518055, China}

\author{Anna Steffinlongo}
\affiliation{Dipartimento di Fisica e Astronomia ``G.Galilei",	Universit\`a degli Studi di Padova, I-35131 Padua, Italy}
\affiliation{Institute for Quantum Optics and Quantum Information - IQOQI Vienna, Austrian Academy of Sciences, Boltzmanngasse 3, 1090 Vienna, Austria}
\affiliation{Atominstitut,  Technische  Universit{\"a}t  Wien, Stadionallee 2, 1020  Vienna,  Austria}

\author{Bixiang Guo}
\affiliation{Shenzhen Institute for Quantum Science and Engineering and Department of Physics, Southern University of Science and Technology, Shenzhen, 518055, China}
\affiliation{Guangdong Provincial Key Laboratory of Quantum Science and Engineering, Southern University of Science and Technology, Shenzhen, 518055, China}

\author{Biyao Liu}
\affiliation{Shenzhen Institute for Quantum Science and Engineering and Department of Physics, Southern University of Science and Technology, Shenzhen, 518055, China}
\affiliation{Guangdong Provincial Key Laboratory of Quantum Science and Engineering, Southern University of Science and Technology, Shenzhen, 518055, China}

\author{Shufeng Xu}
\affiliation{Shenzhen Institute for Quantum Science and Engineering and Department of Physics, Southern University of Science and Technology, Shenzhen, 518055, China}
\affiliation{Guangdong Provincial Key Laboratory of Quantum Science and Engineering, Southern University of Science and Technology, Shenzhen, 518055, China}

\author{Nicolas Gisin}
\email{Nicolas.Gisin@unige.ch}
\affiliation{Group of Applied Physics, University of Geneva, 1211 Geneva 4, Switzerland}
\affiliation{Schaffhausen Institute of Technology - SIT, Geneva, Switzerland}

\author{Armin Tavakoli}
\email{Armin.Tavakoli@oeaw.ac.at}
\affiliation{Institute for Quantum Optics and Quantum Information - IQOQI Vienna, Austrian Academy of Sciences, Boltzmanngasse 3, 1090 Vienna, Austria}
\affiliation{Atominstitut,  Technische  Universit{\"a}t  Wien, Stadionallee 2, 1020  Vienna,  Austria}

\author{Jingyun Fan}
\email{fanjy@sustech.edu.cn}
\affiliation{Shenzhen Institute for Quantum Science and Engineering and Department of Physics, Southern University of Science and Technology, Shenzhen, 518055, China}
\affiliation{Guangdong Provincial Key Laboratory of Quantum Science and Engineering, Southern University of Science and Technology, Shenzhen, 518055, China}

\date{\today}

\begin{abstract}
It has recently been discovered that the nonlocality of an entangled qubit pair can be recycled for several Bell experiments. Here, we go beyond standard Bell scenarios and investigate the recycling of nonlocal resources in a quantum network. We realise a photonic quantum 3-branch star network in which three sources of entangled pairs independently connect three outer parties with a central node. After measuring, each outer party respectively relays their system to an independent secondary measuring party. We experimentally demonstrate that the outer parties can perform unsharp measurements that are strong enough to violate a network Bell inequality with the central party, but weak enough to maintain sufficient entanglement in the network to allow the three secondary parties to do the same. Moreover, the violations are strong enough to exclude any model based on standard projective measurements on the EPR pairs emitted in the network. Our experiment brings together the research program of recycling quantum resources with that of Bell nonlocality in networks.
\end{abstract}

\maketitle

\textit{Introduction.---} Whereas classical systems can be measured without disturbance, quantum measurements influence the state they act on. For instance, a standard dichotomic measurement on one share of an entangled qubit pair causes the entanglement to be lost in the post-measurement state. However, by only weakly coupling the measurement device to the state, one encounters a natural trade-off: the less information extracted, the smaller the induced perturbation in the state \cite{Fuchs1996}. In recent years, much research has been directed at leveraging such trade-offs in order to recycle quantum resources, i.e.~to extract enough information to violate classical limitations, but afterwards preserve enough of the resource to enable further independent quantum information tasks. This includes recycling of quantum communications \cite{Mohan2019, Miklin2020, Kumari2019, Anwer2020, Foletto2020, Anwer2021}, entanglement and steering \cite{Bera2018, Sasmal2018, Coyle2018, Shenoy2019, Choi2020, Pandit2022}, and perhaps most remarkably nonlocality \cite{Gallego2014, Silva2015, Brown2020, Cabello2018, Das2019, Calderaro2020, Cheng2021, Chengb2021}. Using photonic setups, it has been experimentally demonstrated that one share of a singlet state can be recycled to produce two sequential violations of the CHSH inequality \cite{Schiavon2017, Hu2018}. Beyond its conceptual appeal, recycled nonlocality has noteworthy  applications in quantum random number generation \cite{Curchod2017, Bowles2020, Foletto2021} and it constitutes a natural scenario for certifying quantum instruments (see e.g.~\cite{Mohan2019, Miklin2020}).

In parallel with the progress in recycling quantum resources, there has been rapid advances in the development of quantum networks \cite{Stucki2011, Wang2014, Frohlich2013, Liao2018, Chen2021, Hanson2021}. Such networks are not only one of the major promises of quantum technologies \cite{Kimble2008, Wehner2018}, but they also provide conceptual insights into quantum theory (see \cite{ReviewNonloc} for a review). Networks constitute natural generalisations of the traditional Bell scenario: several parties are connected via multiple independent sources that distribute entangled states in some network configuration \cite{Branciard2010, Fritz2012}. They enable interesting new possibilities such as both stronger \cite{Branciard2010, Branciard2012} and novel forms \cite{Tavakoli2021} of entanglement-swapping experiments, nonlocality without inputs \cite{Renou2019}, limitations on measurement dependence \cite{Chaves2021} and  distinguishing the role of complex numbers in quantum theory \cite{Renou2021}. This has motivated several implementations of network nonlocality experiments \cite{Saunders2017, Carvacho2017, Sun2019, Poderini2020, Baumer2020, Agresti2021, Li2022, Chen2022b}.

\begin{figure}[htbp!]
	\centering
	\includegraphics[width=1\linewidth]{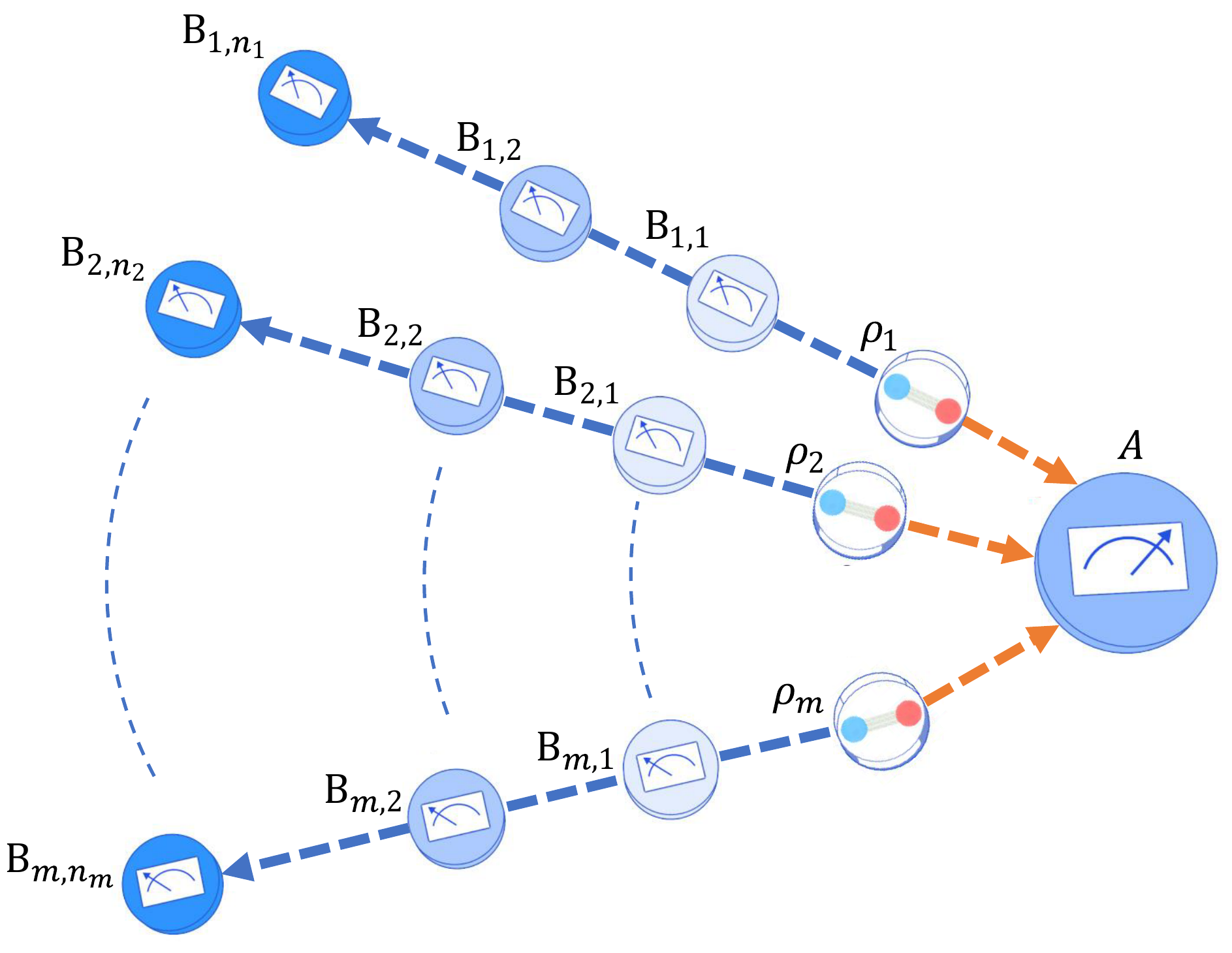}
	\caption{Scenario. Recycling one share of each of the $m$ independent entangled states in a quantum star network with the aim that any selection of $m$ parties (one from each branch) can violate a star network Bell inequality with Alice.}
	\label{Fig1}
\end{figure}

Here, we bring these ideas together by experimentally recycling the network nonlocality enabled by three separate sources of two-qubit entanglement. We consider a star-network configuration \cite{Sende2005, Tavakoli2014, Tavakoli2016} in which a central node is pairwise connected to outer parties via independent sources of bipartite entanglement. By performing suitable measurements on their shares, the parties can together violate a network Bell inequality \cite{Tavakoli2014}. In our scenario, each outer party relays their share to an independent, secondary, outer party with the purpose of enabling another demonstration of network nonlocality (see Figure~\ref{Fig1}). This is based on employing unsharp measurements in the outer parties. Interestingly, although unsharp measurements seem natural for recycling protocols, it has been found that stochastic combinations of standard, projective, measurements is sufficient to recycle violations of the CHSH inequality \cite{Steffinlongo2022}. In our experiment, we demonstrate recycled network nonlocality of a magnitude large enough to exclude the possibility of a simulation based on standard measurements on a maximally entangled state, thus showcasing the genuine usefulness of unsharp measurements.

\textit{Scenario and theory.---} Consider a star network composed of a central node called $A$ who is connected to $m$ branches via independent sources that emit a bipartite quantum state $\rho_k$. The other share of each $\rho_k$ is given to an outer party $B_{k,1}$. Each outer party randomly and privately selects an input $y_{k,1}\in\{0,1\}$, then performs a measurement $\{B^{(k,1)}_{b_{k,1}|y_{k,1}}\}$ on the single share at their disposal, obtaining outcome $b_{k,1}\in\{0,1\}$. Subsequently, the post-measurement state of the $k$'th outer party is relayed to a second, independent, party $B_{k,2}$,  who similarly has an (unbiased) input $y_{k,2}\in\{0,1\}$ and obtains output $b_{k,2}\in\{0,1\}$. This process continues until the final party in the $k$'th branch, $B_{k,n}$, performs a measurement labelled $y_{k,n}\in\{0,1\}$ and records the outcome $b_{k,n}\in\{0,1\}$. Party $A$ also selects an input, $x\in\{0,1\}$, performs a measurement $\{A_{a|x}\}$ of the $m$ shares at its disposal, obtaining an outcome $a\in\{0,1\}$. The scenario is illustrated in Figure~\ref{Fig1}.

A standard test of nonlocality in the network does not involve the relaying of the state, i.e.~it only involves parties $A,B_{1,1},\ldots,B_{m,1}$. For simplicity, we let $s$ be a string of length $m$ with  $s=(1,\ldots,1)$. We can then represent their outcome statistics with the probability distribution $p(a,b_s|x,y_s)$, where $b_s=(b_{1,s_1},\ldots, b_{m,s_m})$ and $y_s=(y_{1,s_1},\ldots,y_{m,s_m})$. In a quantum model, this is given by 
\begin{equation}\label{born}
p(a,b_s|x,y_s)=\Tr\left[\left(\bigotimes_{k=1}^m B^{(k,s_k)}_{b_{k,s_k}|y_{k,s_k}} \otimes A_{a|x}\right) \chi_{s}\right],
\end{equation}
where the ordering of the tensor product has been left implicit. Here, $\chi_{s}$ is the total network state shared between the relevant measuring parties. Thus, in the standard (non-recycling) case of $s=(1,\ldots,1)$, the total state is merely that emitted by all sources, $\chi_{(1,\ldots,1)}=\bigotimes_{k=1}^m \rho_k$.

 A probability distribution is said to admit a network local model if it can be reproduced by associating each source to a local variable $\lambda_k$,
\begin{equation}
p(a,b_s|x,y_s)=\int d\lambda p(a|x,\lambda)\prod_{k=1}^m q_k(\lambda_k)p(b_{k,s_k}|y_{k,s_k},\lambda_k),
\end{equation}
where $\lambda=(\lambda_1,\ldots,\lambda_m)$. The independence of the sources is represented in the fact that each $\lambda_k$ is assigned an independent probability density $q_k(\lambda_k)$. Every network local model satisfies the following network Bell inequality \cite{Tavakoli2014},
\begin{equation}\label{bell}
S_{s}\equiv |I_{s}|^{1/m}+|J_{s}|^{1/m}\leq 1,
\end{equation}
where $I_{s}=\frac{1}{2^m}\sum_{y_s}\sum_{a,b_s} (-1)^{a+\sum_k b_{k,s_k}}p(a,b_s|x=0,y_s)$ and $J_s=\frac{1}{2^m}\sum_{y_s} \sum_{a,b_s} (-1)^{a+\sum_k( b_{k,s_k}+y_{k,s_k})}p(a,b_s|x=1,y_s)$. Quantum models \eqref{born} can achieve the violation $S_s=\sqrt{2}$ \cite{Tavakoli2014}.

Our aim is to recycle the network nonlocality enabled by the $m$ sources in such a way that any choice of $m$ outer parties (one from each branch) can violate the inequality \eqref{bell} together with party $A$. Thus, we require states and measurements in the network such that any selected set of parties $A,B_{1,s_1},B_{2,s_2},\ldots,B_{m,s_m}$, corresponding to some arbitrary $m$-element string $s$, obtains a probability distribution $p(a,b_{s}|x,y_s)$ for which $S_s>1$. This requires every party in the $k$'th branch (except the final one) to preserve some of the resource $\rho_k$ after their measurements, such that the subsequent parties in the branch can also demonstrate a violation of \eqref{bell}. For any chosen set of parties $s$, the relevant probability distribution is given by \eqref{born} but now based on the recycled total state on average, $\chi_s$. This state is obtained by  recursively  applying the L\"uders rule to each state originally created in the network,
\begin{align}\nonumber
\rho_k^{(j)}=\frac{1}{2}\sum_{b_{k,j},y_{k,j}}\left(\sqrt{B^{(k,j)}_{b_{k,j}|y_{k,j}}}\otimes \openone  \right)\rho_k^{(j-1)} \left(\sqrt{B^{(k,j)}_{b_{k,j}|y_{k,j}}}\otimes \openone  \right),
\end{align}
where $\rho_k^{(j)}$ ($\rho_k^{(0)}=\rho_k$) is the average state associated to source $k$ after the measurements of parties $B_{k,1},\ldots,B_{k,j}$. We then have $\chi_s=\bigotimes_{k=1}^m \rho_k^{(s_k-1)}$.

\begin{figure*}[htbp!]
	\centering
	\includegraphics[width=\linewidth]{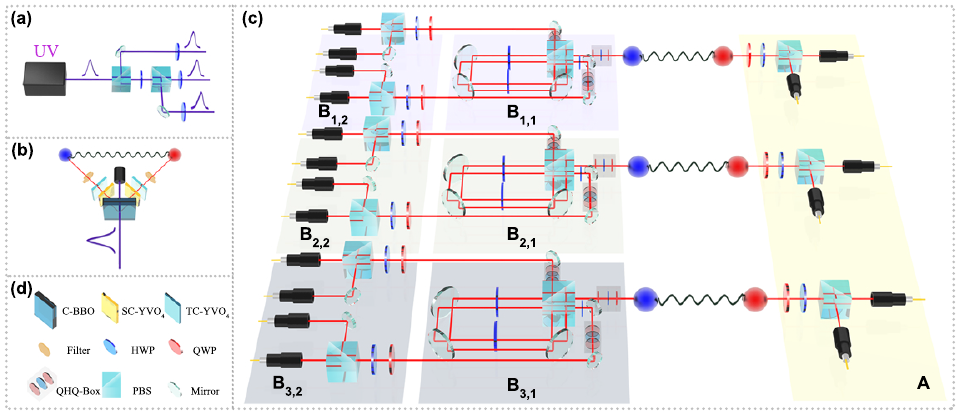}
	\caption{Experimental schematic of recycling nonlocality in a quantum network. 
		(a) We split a UV laser pulse with pulse duration of $\tau=150$ fs and $\lambda=390$ nm into three pulses of equal energy. (b) Each pulse induces SPDC in a pair of BBO crystals (with a HWP in-between) to probabilistically emit a pair of photons at $\lambda=780$ nm in Bell state $\ket{\Phi^+}$~\cite{Kwait1995PRL,Zhang2015PRL,Wang2016PRL} as the independent source. Pairs of SC-YVO4 and TC-YVO4 crystals are used to remove spatial and temporal walk-offs due to orthogonally polarized photons traversing BBO. (c) Party A is connected to three branches via three sources. In each branch, party $B_{k,1}$ performs an unsharp measurement to the photon and then relays the photon to party $B_{k,2}$. The unsharp measurement is implemented in a Sagnac interferometer, in which the photon in polarization states $\ket{H}$ and $\ket{V}$ travels along separate spatial paths. We use a HWP to vary the coupling between the polarization degree-of-freedom and the path degree-of-freedom to adjust unsharp measurement $\eta^{X}_{k,1}\sigma_X$ or $\eta^{Z}_{k,1}\sigma_Z$. We use QHQ-Box consisting of two quarter-wave plate (QWP) and a HWP to set a photon to any polarization state on Bloch sphere. Party $B_{k,2}$ uses a HWP, a QWP and a polarizing beamsplitters (PBS) to perform projective polarization measurement to single photons and send them to detectors via optical fiber. All photon detection events are time-tagged for correlation analysis (not shown).  
	}
	\label{fig:FIG2}
\end{figure*}

Consider now the following quantum protocol. Each source distributes the maximally entangled state $\ket{\phi^+}=\frac{1}{\sqrt{2}}\left[\ket{00}+\ket{11}\right]$. All outer parties perform unsharp measurements in the Pauli $\sigma_Z$ and $\sigma_X$ bases respectively. The $j$'th one in the $k$'th branch has observables 
$\eta^Z_{k,j}\sigma_Z$ 
($y_{k,j}=0$) and $\eta^X_{k,j}\sigma_X$ 
($y_{k,j}=1$).  
Let $A$ perform measurements $A_{a|x}=\sum_{a_1\oplus \ldots\oplus a_m=a} \bigotimes_{k=1}^m \Pi^{(k)}_{a_k|x}$, where $ \Pi^{(k)}_{a_k|x}=\frac{1}{2}\left(\openone+(-1)^{a_k}\left(\cos\theta\sigma_Z+(-1)^x\sin\theta\sigma_X\right)\right)$. This measurement can be interpreted as $m$ separate sharp single-qubit measurements followed by a parity wiring to determine the final outcome. For any choice of parties, $s$, for which the recycling is considered, one finds (see Supplemental Material)
\begin{align}\label{eq}
	& S_{s}=\prod_{k=1}^{m}2^{\frac{1-s_k}{m}} \left(\eta^Z_{k,s_k}\cos\theta \prod_{j=1}^{s_k-1}f^X_{k,j}+ \eta^X_{k,s_k}\sin\theta \prod_{j=1}^{s_k-1}f^Z_{k,j}\right)^{\frac{1}{m}},
\end{align}
where $f^u_{k,j}=1+\sqrt{1-\left(\eta^u_{k,j}\right)^2}$ for $u\in\{Z,X\}$. Such a strategy is sufficient to demonstrate recycling of network nonlocality for any choice of parties, $s$, for any number of sequential measurements per branch, $n$. This follows immediately from the fact that a longer sequence (in any branch) cannot improve the value of $S_s$. If we were to consider the worst case, in which all elements of $s$ are reset to the largest element in $s$, then \eqref{eq} reduces to the correlation quantity obtained for recycling violations of the CHSH inequality (choose $\eta_{k,j}=\eta_j$), for which it was shown that appropriate choices of  $\{\eta^Z_{k,j},\eta^X_{k,j}\}$ and a $\theta$ exist ~\cite{Brown2020}.

\textit{Experiment.---}
We conduct a proof-of-principle experimental demonstration following the protocol introduced in the above. As shown in Figure.~\ref{fig:FIG2}, party $A$ is connected to three branches ($m=3$) via independent sources and each branch has two parties ($B_{k,1}$ and $B_{k,2}$).
We split a UV laser pulse with wavelength $\lambda=390$ nm into three parts and focus each of them on a pair of $\beta$-barium borate crystals (BBOs). The induced spontaneous parametric down-conversion (SPDC) process emits probabilistically a pair of photons at $\lambda=780$ nm~\cite{Kwait1995PRL,Zhang2015PRL,Wang2016PRL}. We pass the generated photons through 3-nm bandpass filters and keep the generation probability of photon pairs in SPDC as low as 0.02 per pulse to mitigate multi-photon effect. We estimate that the state fidelity of the pairs of photons with respect to the targeted Bell state $\ket{\Phi^+}=(\ket{HH}+\ket{VV})/\sqrt{2}$ is greater than 0.96~\cite{James2001PRATomo}, where $\ket{H}$ and $\ket{V}$ respectively represent the horizontal and vertical polarization states. 

\begin{figure*}
	\centering
	\includegraphics[width=\linewidth]{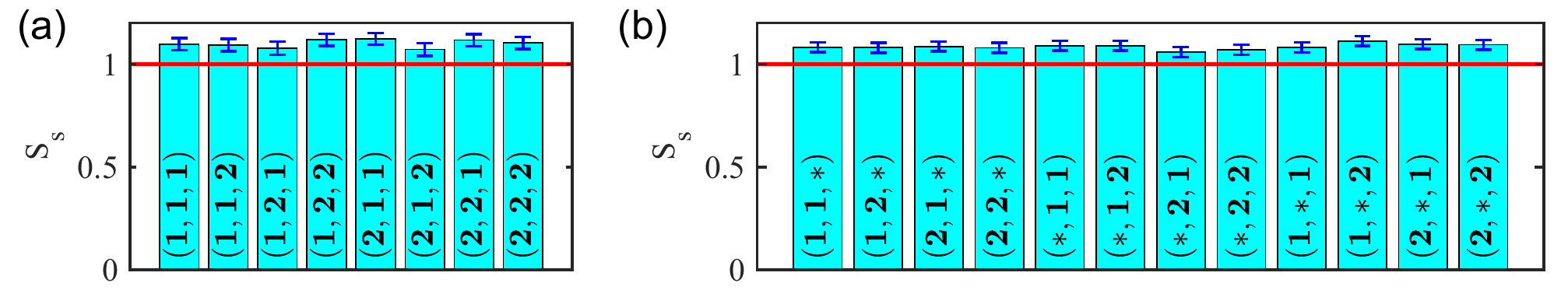}
	\caption{Simultaneous experimental violations of network Bell inequality (\ref{bell}) with any choice of three or two outer parties from separate branches together with party $A$ in (a) and (b), respectively, where $*$ represents the branch unused in the respective study. Error bars shown in (a)(b) represent one standard deviation in the experiment.} 
	\label{fig:FIG3}
\end{figure*}

 In each branch, one photon from the source is sent to party $A$, the other photon is sent to outer party $B_{k,1}$. $B_{k,1}$ performs unsharp measurement to the photon via a Controlled-Not (CNOT) gate~\cite{CNOT1,CNOT2,CNOT3}. In this proof-of-principle experiment, $B_{k,1}$ encodes system qubit ($\ket{\cdot}$) to the polarization state and meter qubit ($\ket{\cdot}_m$) to the path state of the photon at his disposal. As shown in Fig.~\ref{fig:FIG2}, vertically (horizontally) polarized photons in system state $\ket{0}$ ($\ket{1}$) propagate (counter-) clockwisely upon incidence on the Sagnac interferometer, hence coupling to the meter qubit $\ket{0}_m$ ($\ket{1}_m$), respectively. We vary the half-wave plate (HWP) in the Sagnac interferometer to adjust the unsharp measurement strength, i.e., the system-meter coupling strength, $\{\eta^Z_{k,j},\eta^X_{k,j}\}\in[0,1]$~\cite{Maoyali2019PRL}. After exiting the Sagnac interferometer, photons taking the upper (lower) path, which corresponds to measurement outcome $b_{k,1}=0$ ($1$), are relayed to party $B_{k,2}$ for projectve polarization measurements. Finally, the photon detection event reported by one of the four single-photon detectors per pulse corresponds to exactly one of the four outcome combinations ($b_{k,1}$, $b_{k,2})$=$(0,0),(0,1),(1,0),(1,1)$ in this branch, respectively. (see Supplemental Material for details). 
 
 We record six-photon coincidence events within a time window of 5 ns, with three photons detected by party $A$ and one photon detected in each branch, at a rate of $0.7~\text{s}^{-1}$ with an experimental repetition rate of 80 MHz. Consider that the optical path length of UV pulses traveling to BBO crystals in free space varies due to the environmental instability such as the variation in ambient temperature. The phase of the UV laser pulse is randomized at a rate of a few hundred hertz or higher, effectively making three sources mutually independent~\cite{Saunders2017}.

\begin{figure}[htbp!]
	\centering
	\includegraphics[width=1\linewidth]{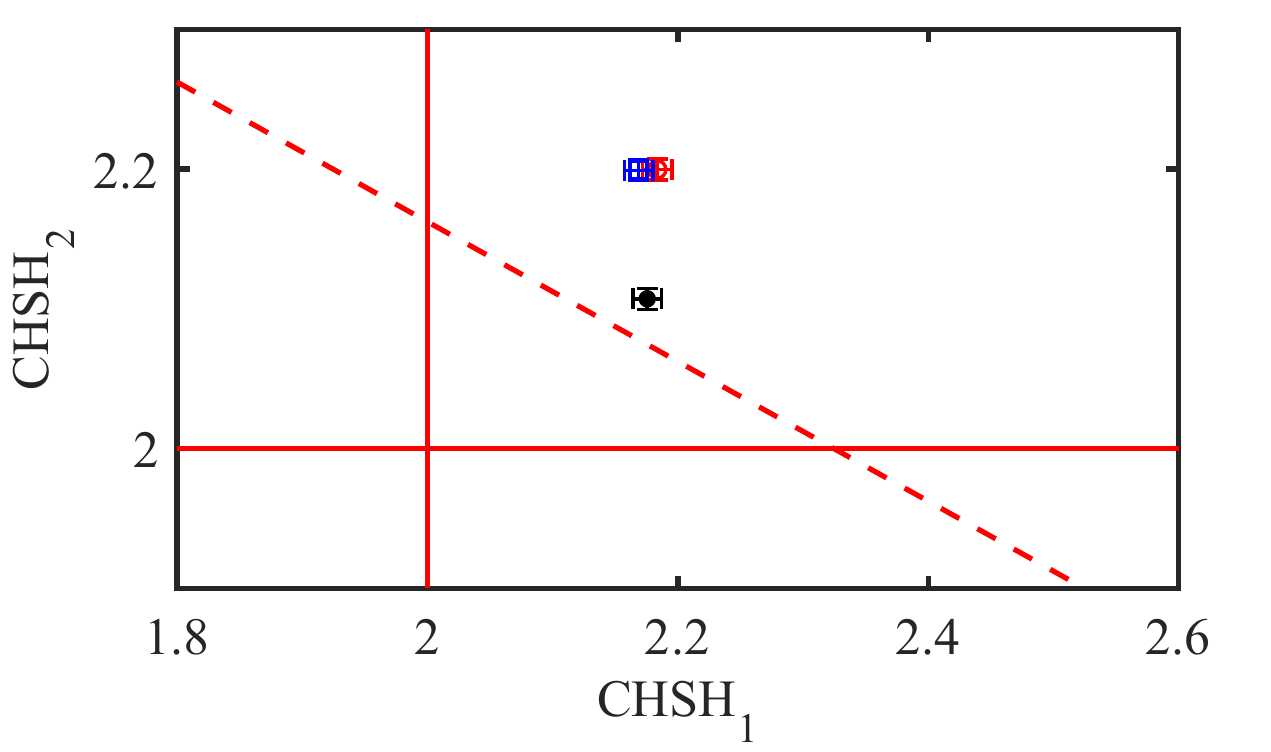}
	\caption{Experimental sequential test of CHSH-Bell inequality in each branch ($k=1,2,3$). The CHSH$_1$ (CHSH$_2$) is measured between $A$ and $B_{k,1}$ ($B_{k,2}$). The three measured double-values are  ($2.1834\pm0.0116$, $2.1997\pm0.0074$), ($2.1755\pm0.0114$, $2.1072\pm0.0075$), ($2.1685\pm0.0114$, $2.1991\pm0.0073$), illustrated by the open dot, filled dot and open square respectively. Error bars represent one standard deviation. The dashed line represents the optimal trade-off based on projective measurements and shared randomness. The solid line is the local model bound.}
	\label{fig:FIG4}
\end{figure}

By appropriately choosing the parameter values, any choice of three outer parties, one from each branch, can simultaneously violate the network Bell inequality \eqref{bell} together with party $A$. Fig.~\ref{fig:FIG3}(a) presents such an experimental demonstration with $\eta^Z_{j=1}=\eta^X_{j=1}=0.8$, $\eta^Z_{j=2}=\eta^X_{j=2}=1$ independent of the choice of $k$, and $\theta=45^\circ$ (which corresponds to unbiased bases). In this so-called 3-local scenario, we collected 5550 six-photon coincidence events in total over 64 measurement settings, with which we obtained an average value of $S_{avg}=\frac{1}{8}\Sigma_s S_s =1.100\pm0.030$ (taken over all 8 strings $s$, with the worst-case value $S_{(2,1,2)}=1.071\pm0.032$) surpassing the limit set by network local models by more than three standard deviations. With the same set of experimental data, we show in Fig.~\ref{fig:FIG3}(b) that any choice of two outer parties from separate branches, namely in the bi-local scenario, simutaneously violate the network Bell inequality \eqref{bell} together with party $A$, for which we obtain an average value of  $S_{avg}=\frac{1}{12}\Sigma_s S_s =1.085\pm0.024$ surpassing the limit set by network local models by more than three standard deviations.

Unsharp measurements in the branches are the central component of the experiment. We now show that this resource is indispensible for the experimental results, i.e.~that the results cannot be reproduced in a simpler setup, based on projective measurements on maximally entangled qubit states. To this end, we note that by keeping the outcomes $(a_1,a_2,a_3)$ of Alice, instead of wiring them into $a$, the experimental data can also be used for three separate (one per branch) sequential tests of CHSH-Bell inequality~\cite{PhysRevLett.23.880}, i.e.,~between party $A$, and $B_{k,1}$ and $B_{k,2}$ respectively, for $k=1,2,3$. In a given branch, the two CHSH parameters are $\text{CHSH}^{(k)}_j=\sum_{x,y_{k,j}}(-1)^{xy_{k,j}}\expect{A_x,B_{k,j}}_{\rho_k^{(j-1)}}$, 
for $j\in\{1,2\}$. In local models, the values exhibit no trade-off, each being limited by the local bound, i.e.~$(\text{CHSH}_1,\text{CHSH}_2)\leq (2,2)$. However, a quantum double violation, i.e.~$(\text{CHSH}_1,\text{CHSH}_2)>(2,2)$, does not rule out an implementation based on projective measurements. It has been shown that projective qubit measurements on a maximally entangled state, assisted by classical shared randomness, can achieve at most the double-violation trade-off $\text{CHSH}_2=\sqrt{10}-\frac{\text{CHSH}_1}{2}$, for the relevant range $\text{CHSH}_1\in\left[2,2\sqrt{10}-4\right]$. Thus, to certify the need for unsharp measurements and to justify the use of this experimentally more complex resource, we investigate whether our sequential CHSH values exceed the limitation of projective strategies. The three experimentally measured pairs of CHSH values are illustrated in Figure~\ref{fig:FIG4}. We find statistically significant violations of the projective limit in all three branches. This demonstrates the need for unsharp measurements in our network scenario. 

\textit{Conclusions.---}  We have shown that the network nonlocality originally enabled by multiple independent sources can be recycled arbitrarily many times by parties arranged in a star network. We experimentally demonstrated this in a network based on three sources of polarisation qubit entanglement and three instances of unsharp measurements, allowing for sufficient preservation of entanglement to enable a second violation. Our reported double violations are sufficiently strong to exclude the possibility of simulating the experiment using only projective measurements.

The network Bell inequalities that we have considered are closely related to the CHSH inequality for standard Bell scenarios \cite{Andreoli2017, Tavakoli2017} and they do not require entangled measurements in the center node. A natural next step is therefore to investigate recycling of network nonlocality based on entanglement swapping. Although such scenarios can still maintain conceptual links to standard Bell scenarios \cite{Gisin2017}, there has recently been several proposals of entanglement swapping experiments that constintute more genuine notions of network nonlocality \cite{Tavakoli2021, Pozas2022, Supic2021}. A particularly interesting prospect is to investigate unsharp implementations of joint, entangled, measurements for recycling network nonlocality.

\textit{Note added.---} While completing this manuscript, we became aware of Ref.~\cite{Hou2021} which studies recycling violations of bilocality.

\section*{Acknowledgements}
Y.-L.M., Z.-D.L and J.F. are supported by the Key-Area Research and Development Program of Guangdong Province Grant No.2020B0303010001, Grant No.2019ZT08X324, Guangdong Provincial Key Laboratory Grant No.2019B121203002, and the National Natural Science Foundation of China Grants No.12004207 and No.12005090. A.T. is supported by the Wenner-Gren Foundations. N.G. is supported by the Swiss NCCR SwissMap.

\bibliography{references_sequentialnetwork_070222}
 
 \clearpage
 
 \onecolumngrid
 \subsection*{\textbf{\large Supplementary Materials for Recycling nonlocality in a quantum network}}

\section{Derivation of Eq. (4) in main text}
Recall that for a choice of $m$ parties, $s$, the quantum probabilities are given by
\begin{equation}\label{born}
	p(a,b_s|x,y_s)=\Tr\left[\left(\bigotimes_{k=1}^m B^{(k,s_k)}_{b_{k,s_k}|y_{k,s_k}} \otimes A_{a|x}\right) \chi_{s}\right],
\end{equation}
up to the ordering of the tensor product. The recycled total state, $\chi_s$, takes the form $\chi_s=\bigotimes_{k=1}^m \rho_k^{(s_k-1)}$, where $\rho_k^{0}=\rho_k$. Here, $\rho_k^{(j-1)}$ is the average recycled state that arrives to $B_{k,j}$. These states can be obtained  recursively from the L\"uders rule,
\begin{align}
	\rho_k^{(j)}=\frac{1}{2}\sum_{b_{k,j},y_{k,j}}\left(\sqrt{B^{(k,j)}_{b_{k,j}|y_{k,j}}} \otimes \openone \right)\rho_k^{(j-1)} \left(\sqrt{B^{(k,j)}_{b_{k,j}|y_{k,j}}} \otimes \openone \right),
\end{align}

In the quantum protocol, we let each source distribute the maximally entangled state, $\ket{\phi^+}=\frac{1}{\sqrt{2}}\left[\ket{00}+\ket{11}\right]$. All outer parties perform unsharp measurements in the Pauli Z and X bases respectively. The $j$'th one in the $k$'th branch has observables $\eta^Z_{k,j}\sigma_Z$($y_{k,j}=0$) and $\eta^X_{k,j}\sigma_X$($y_{k,j}=1$). Let $A$ perform measurements $A_{a|x}=\sum_{a_1\oplus \ldots\oplus a_m=a} \bigotimes_{k=1}^m \Pi^{(k)}_{a_k|x}$, where $\Pi^{(k)}_{a_k|x}=\frac{1}{2}\left(\openone+(-1)^{a_k}\left(\cos\theta\sigma_Z+(-1)^x\sin\theta\sigma_X\right)\right)$.  We now recursively compute the recycled states.
\begin{equation}
	\begin{aligned}
		&\rho_k^{(j)}=\frac{1}{2}\sum_{b_{k,j},y_{k,j}}\left(\sqrt{B^{(k,j)}_{b_{k,j}|y_{k,j}}}\otimes \openone\right)\rho_k^{(j-1)}\left(\sqrt{B^{(k,j)}_{b_{k,j}|y_{k,j}}}\otimes \openone\right)\\
		&=\frac{1}{2}\left[\sqrt{\frac{1}{2}(\openone+\eta^Z_{k,j}\sigma_{{Z}})}\otimes \openone\right]\rho_k^{(j-1)}\left[\sqrt{\frac{1}{2}(\openone+\eta^Z_{k,j}\sigma_{{Z}})}\otimes\openone\right]+\frac{1}{2}\left[\sqrt{\frac{1}{2}(\openone-\eta^Z_{k,j}\sigma_{{Z}})}\otimes \openone \right]\rho_k^{(j-1)}\left[\sqrt{\frac{1}{2}(\openone-\eta^Z_{k,j}\sigma_{{Z}})}\otimes\openone \right]\\	
		&\left.+\frac{1}{2}\left[\sqrt{\frac{1}{2}(\openone+\eta^X_{k,j}\sigma_{{X}})} \otimes \openone \right]\rho_k^{(j-1)}\left[\sqrt{\frac{1}{2}(\openone+\eta^X_{k,j}\sigma_{{X}})} \otimes \openone \right]+\frac{1}{2}\left[\sqrt{\frac{1}{2}(\openone-\eta^X_{k,j}\sigma_{{X}})} \otimes \openone \right]\rho_k^{(j-1)}\left[\sqrt{\frac{1}{2}(\openone-\eta^X_{k,j}\sigma_{{X}})} \otimes \openone\right]\right.\\
		&=\frac{1}{4}\left[2+\sqrt{1-\left(\eta^Z_{k,j}\right)^2}+\sqrt{1-\left(\eta^X_{k,j}\right)^2}\right]\rho_k^{(j-1)}+\frac{1}{4}\left[1-\sqrt{1-\left(\eta^Z_{k,j}\right)^2}\right](\sigma_{Z}\otimes\openone)\rho_k^{(j-1)}(\sigma_{{Z}}\otimes\openone)\\&+\frac{1}{4}\left[1-\sqrt{1-\left(\eta^X_{k,j}\right)^2}\right](\sigma_{{X}} \otimes \openone)\rho_k^{(j-1)}(\sigma_{{X}}\otimes \openone).\\	
		\label{RhoABk}
	\end{aligned}
\end{equation}

To compute the network Bell inequality, we must evaluate the quantities 
\begin{align}
	&I_{s}=\frac{1}{2^m}\sum_{y_s}\sum_{a,b_s} (-1)^{a+\sum_k b_{k,s_k}}p(a,b_s|0,y_s),\\
	& J_s=\frac{1}{2^m}\sum_{y_s} \sum_{a,b_s} (-1)^{a+\sum_k(b_{k,s_k}+y_{k,s_k}) }p(a,b_s|1,y_s).
\end{align}
We now compute these quantities in the quantum model. 
\begin{equation}
	\begin{aligned}
		&I_{s}=\frac{1}{2^m}\sum_{y_s}\sum_{a,b_s} (-1)^{a+\sum_k b_{k,s_k}}p(a,b_s|0,y_s)\\
		&=\frac{1}{2^m}\sum_{y_s}\sum_{a,b_s} (-1)^{a+\sum_k b_{k,s_k}}\Tr\left[\left(\bigotimes_{k=1}^m B^{(k,s_k)}_{b_{k,s_k}|y_{k,s_k}} \otimes A_{a|0}\right) \chi_{s}\right]\\	
		&=\frac{1}{2^m}\sum_{y_s}\sum_{a,b_s} (-1)^{a+\sum_k b_{k,s_k}}\Tr\left[\left(\bigotimes_{k=1}^m B^{(k,s_k)}_{b_{k,s_k}|y_{k,s_k}} \otimes \sum_{a_1\oplus \ldots\oplus a_m=a} \bigotimes_{k=1}^m \Pi^{(k)}_{a_k|0}\right) \bigotimes_{k=1}^m \rho_k^{(s_k-1)}\right]\\
		&=\frac{1}{2^m}\sum_{y_s}\sum_{a,b_s}\sum_{a_1\oplus \ldots\oplus a_m=a} (-1)^{a+\sum_k b_{k,s_k}}\Tr\left[\bigotimes_{k=1}^m \left(B^{(k,s_k)}_{b_{k,s_k}|y_{k,s_k}} \otimes\Pi^{(k)}_{a_k|0}\rho_k^{(s_k-1)}\right)\right]\\
		&=\frac{1}{2^m}\sum_{y_s}\sum_{a,b_s}\sum_{a_1\oplus \ldots\oplus a_m=a} (-1)^{a+\sum_k b_{k,s_k}}\prod_{k=1}^m \left[\Tr\left(B^{(k,s_k)}_{b_{k,s_k}|y_{k,s_k}} \otimes\Pi^{(k)}_{a_k|0}\rho_k^{(s_k-1)}\right)\right]\\	
		&=\frac{1}{2^m}\prod_{k=1}^{m}\Tr\left\{\rho_k^{(s_k-1)}\left[(B^{(k,s_k)}_{0|0}-B^{(k,s_k)}_{1|0})\otimes\Pi^{0}_{A_m}+(B^{(k,s_k)}_{0|1}-B^{(k,s_k)}_{1|1})\otimes\Pi^{0}_{A_m}\right]\right\}\\
		&=\frac{1}{2^m}\prod_{k=1}^{m}\left\{\eta^Z_{k,s_k}\Tr\left[\rho_k^{(s_k-1)}(\cos\theta\sigma_{{Z}}\otimes\sigma_{{Z}}+\sin\theta\sigma_{{Z}}\otimes\sigma_{{X}})\right]+\eta^X_{k,s_k}\Tr\left[\rho_k^{(s_k-1)}(\cos\theta\sigma_{{X}}\otimes\sigma_{{Z}}+\sin\theta\sigma_{{X}}\otimes\sigma_{{X}})\right]\right\}\\
		&=\frac{1}{2^m}\prod_{k=1}^{m}\left\{\eta^Z_{k,s_k}\cos\theta \Tr\left[\rho_k^{(s_k-1)}(\sigma_{{Z}}\otimes\sigma_{{Z}})\right]+\eta^Z_{k,s_k}\sin\theta \Tr\left[\rho_k^{(s_k-1)}(\sigma_{{Z}}\otimes\sigma_{{X}})\right]\right.\\&\left.+\eta^X_{k,s_k}\cos\theta \Tr\left[\rho_k^{(s_k-1)}(\sigma_{{X}}\otimes\sigma_{{Z}})\right]+\eta^X_{k,s_k}\sin\theta \Tr\left[\rho_k^{(s_k-1)}(\sigma_{{X}}\otimes\sigma_{{X}})\right]\right\}
		\label{IK}
	\end{aligned}
\end{equation}

Below we compute quantities $\Tr\left[\rho_k^{(s_k-1)}(\sigma_{{Z}}\otimes\sigma_{{Z}})\right]$, $\Tr\left[\rho_k^{(s_k-1)}(\sigma_{{Z}}\otimes\sigma_{{X}})\right]$, $\Tr\left[\rho_k^{(s_k-1)}(\sigma_{{X}}\otimes\sigma_{{Z}})\right]$, and $\Tr\left[\rho_k^{(s_k-1)}(\sigma_{{X}}\otimes\sigma_{{X}})\right]$, 
\begin{equation}
	\begin{aligned}
		&\Tr\left[\rho_k^{(s_k-1)}(\sigma_{{Z}}\otimes\sigma_{{Z}})\right]\\
		&=\frac{1}{4}\left[2+\sqrt{1-\left(\eta^Z_{k,s_k-1}\right)^2}+\sqrt{1-\left(\eta^X_{k,s_k-1}\right)^2}\right]\Tr\left[\rho_k^{(s_k-2)}(\sigma_{{Z}}\otimes\sigma_{{Z}})\right]\\&+\frac{1}{4}\left[1-\sqrt{1-\left(\eta^Z_{k,s_k-1}\right)^2}\right]\Tr\left[\rho_k^{(s_k-2)}(\sigma_{{Z}}\sigma_{{Z}}\sigma_{{Z}}\otimes\sigma_{{Z}})\right]+\frac{1}{4}\left[1-\sqrt{1-\left({\eta^X_{k,s_k-1}}\right)^2}\right]\Tr\left[\rho_k^{(s_k-2)}(\sigma_{{X}}\sigma_{{Z}}\sigma_{{X}}\otimes\sigma_{{Z}})\right]\\
		&=\frac{1}{4}\left[2+\sqrt{1-\left({\eta^Z_{k,s_k-1}}\right)^2}+\sqrt{1-\left({\eta^X_{k,s_k-1}}\right)^2}\right]\Tr\left[\rho_k^{(s_k-2)}(\sigma_{{Z}}\otimes\sigma_{{Z}})\right]+\frac{1}{4}\left[1-\sqrt{1-\left(\eta^Z_{k,s_k-1}\right)^2}\right]\Tr\left[\rho_k^{(s_k-2)}(\sigma_{{Z}}\otimes\sigma_{{Z}})\right]\\
		&-\frac{1}{4}\left[1-\sqrt{1-\left(\eta^X_{k,s_k-1}\right)^2}\right]\Tr\left[\rho_k^{(s_k-2)}(\sigma_{{Z}}\otimes\sigma_{{Z}})\right]\\
		&=\frac{1}{2}\left[1+\sqrt{1-\left(\eta^X_{k,s_k-1}\right)^2}\right]\Tr\left[\rho_k^{(s_k-2)}(\sigma_{{Z}}\otimes\sigma_{{Z}})\right].
	\end{aligned}
\end{equation}
\begin{equation}
	\begin{aligned}
		&\Tr\left[\rho_k^{(s_k-1)}(\sigma_{{Z}}\otimes\sigma_{{X}})\right]\\&=\frac{1}{4}\left[2+\sqrt{1-\left({\eta^Z_{k,s_k-1}}\right)^2}+\sqrt{1-\left
			({\eta^X_{k,s_k-1}}\right)^2}\right]\Tr\left[\rho_k^{(s_k-2)}(\sigma_{{Z}}\otimes\sigma_{{X}})\right]\\
		&+\frac{1}{4}\left[1-\sqrt{1-\left({\eta^Z_{k,s_k-1}}\right)^2}\right]\Tr\left[\rho_k^{(s_k-2)}(\sigma_{{Z}}\sigma_{{Z}}\sigma_{{Z}}\otimes\sigma_{{X}})\right]
		+\frac{1}{4}\left[1-\sqrt{1-\left({\eta^X_{k,s_k-1}}\right)^2}\right]\Tr\left[\rho_k^{(s_k-2)}(\sigma_{{X}}\sigma_{{Z}}\sigma_{{X}}\otimes\sigma_{{X}})\right]\\
		&=\frac{1}{4}\left[2+\sqrt{1-\left({\eta^Z_{k,s_k-1}}\right)^2}+\sqrt{1-\left({\eta^X_{k,s_k-1}}\right)^2}\right]\Tr\left[\rho_k^{(s_k-2)}(\sigma_{{Z}}\otimes\sigma_{{X}})\right]+\frac{1}{4}\left[1-\sqrt{1-\left({\eta^Z_{k,s_k-1}}\right)^2}\right]\Tr\left[\rho_k^{(s_k-2)}(\sigma_{{Z}}\otimes\sigma_{{X}})\right]\\
		&-\frac{1}{4}\left[1-\sqrt{1-\left({\eta^X_{k,s_k-1}}\right)^2}\right]\Tr\left[\rho_k^{(s_k-2)}(\sigma_{{Z}}\otimes\sigma_{{X}})\right]\\
		&=\frac{1}{2}\left[1+\sqrt{1-\left({\eta^X_{k,s_k-1}}\right)^2}\right]\Tr\left[\rho_k^{(s_k-2)}(\sigma_{{Z}}\otimes\sigma_{{X}})\right]\\
	\end{aligned}
\end{equation}
\begin{equation}
	\begin{aligned}
		&\Tr\left[\rho_k^{(s_k-1)}(\sigma_{{X}}\otimes\sigma_{{Z}})\right]\\
		&=\frac{1}{4}\left[2+\sqrt{1-\left({\eta^Z_{k,s_k-1}}\right)^2}+\sqrt{1-\left({\eta^X_{k,s_k-1}}\right)^2}\right]\Tr\left[\rho_k^{(s_k-2)}(\sigma_{{X}}\otimes\sigma_{{Z}})\right]\\&+\frac{1}{4}\left[1-\sqrt{1-\left({\eta^Z_{k,s_k-1}}\right)^2}\right]\Tr\left[\rho_k^{(s_k-2)}(\sigma_{{Z}}\sigma_{{X}}\sigma_{{Z}}\otimes\sigma_{{Z}})\right]+\frac{1}{4}\left[1-\sqrt{1-\left({\eta^X_{k,s_k-1}}\right)^2}\right]\Tr\left[\rho_k^{(s_k-2)}(\sigma_{{X}}\sigma_{{X}}\sigma_{{X}}\otimes\sigma_{{Z}})\right]\\ 
		&=\frac{1}{4}\left[2+\sqrt{1-\left({\eta^Z_{k,s_k-1}}\right)^2}+\sqrt{1-\left({\eta^X_{k,s_k-1}}\right)^2}\right]\Tr\left[\rho_k^{(s_k-2)}(\sigma_{{X}}\otimes\sigma_{{Z}})\right]-\frac{1}{4}\left[1-\sqrt{1-\left({\eta^Z_{k,s_k-1}}\right)^2}\right]\Tr\left[\rho_k^{(s_k-2)}(\sigma_{{X}}\otimes\sigma_{{Z}})\right]\\
		&+\frac{1}{4}\left[1-\sqrt{1-\left({\eta^X_{k,s_k-1}}\right)^2}\right]\Tr\left[\rho_k^{(s_k-2)}(\sigma_{{X}}\otimes\sigma_{{Z}})\right]\\
		&=\frac{1}{2}\left[1+\sqrt{1-\left({\eta^Z_{k,s_k-1}}\right)^2}\right]\Tr\left[\rho_k^{(s_k-2)}(\sigma_{{X}}\otimes\sigma_{{Z}})\right].
	\end{aligned}
\end{equation}

\begin{equation}
	\begin{aligned}
		&\Tr\left[\rho_k^{(s_k-1)}(\sigma_{{X}}\otimes\sigma_{{X}})\right]\\
		&=\frac{1}{4}\left[2+\sqrt{1-\left({\eta^Z_{k,s_k-1}}\right)^2}+\sqrt{1-\left({\eta^X_{k,s_k-1}}\right)^2}\right]\Tr\left[\rho_k^{(s_k-2)}(\sigma_{{X}}\otimes\sigma_{{X}})\right]
		\\&+\frac{1}{4}\left[1-\sqrt{1-\left({\eta^Z_{k,s_k-1}}\right)^2}\right]\Tr\left[\rho_k^{(s_k-2)}(\sigma_{{Z}}\sigma_{{X}}\sigma_{{Z}}\otimes\sigma_{{X}})\right]+\frac{1}{4}\left[1-\sqrt{1-\left({\eta^X_{k,s_k-1}}\right)^2}\right]\Tr\left[\rho_k^{(s_k-2)}(\sigma_{{X}}\sigma_{{X}}\sigma_{{X}}\otimes\sigma_{{X}})\right]\\	&=\frac{1}{4}\left[2+\sqrt{1-\left({\eta^Z_{k,s_k-1}}\right)^2}+\sqrt{1-\left({\eta^X_{k,s_k-1}}\right)^2}\right]\Tr\left[\rho_k^{(s_k-2)}(\sigma_{{X}}\otimes\sigma_{{X}})\right]-\frac{1}{4}\left[1-\sqrt{1-\left({\eta^Z_{k,s_k-1}}\right)^2}\right]\Tr\left[\rho_k^{(s_k-2)}(\sigma_{{X}}\otimes\sigma_{{X}})\right]\\
		&+\frac{1}{4}\left[1-\sqrt{1-\left({\eta^X_{k,s_k-1}}\right)^2}\right]\Tr\left[\rho_k^{(s_k-2)}(\sigma_{{X}}\otimes\sigma_{{X}})\right]\\
		&=\frac{1}{2}\left[1+\sqrt{1-\left({\eta^Z_{k,s_k-1}}\right)^2}\right]\Tr\left[\rho_k^{(s_k-2)}(\sigma_{{X}}\otimes\sigma_{{X}})\right]\\
	\end{aligned}
\end{equation}
Recursively, we have 
\begin{equation}
	\begin{aligned}
		\Tr\left[\rho_k^{(s_k-1)}(\sigma_{{Z}}\otimes\sigma_{{Z}})\right]&=2^{1-s_k}\Tr\left[\rho_k^{0}(\sigma_{{Z}}\otimes\sigma_{{Z}})\right]\prod_{j=1}^{s_k-1}\left[1+\sqrt{1-\left({\eta^X_{k,j}}\right)^2}\right],\\
		\Tr\left[\rho_k^{(s_k-1)}(\sigma_{{Z}}\otimes\sigma_{{X}})\right]&=2^{1-s_k}\Tr\left[\rho_k^{0}(\sigma_{{Z}}\otimes\sigma_{{X}})\right]\prod_{j=1}^{s_k-1}\left[1+\sqrt{1-\left({\eta^X_{k,j}}\right)^2}\right],\\
		\Tr\left[\rho_k^{(s_k-1)}(\sigma_{{X}}\otimes\sigma_{{Z}})\right]&=2^{1-s_k}\Tr\left[\rho_k^{0}(\sigma_{{X}}\otimes\sigma_{{Z}})\right]\prod_{j=1}^{s_k-1}\left[1+\sqrt{1-\left({\eta^Z_{k,j}}\right)^2}\right],\\
		\Tr\left[\rho_k^{(s_k-1)}(\sigma_{{X}}\otimes\sigma_{{X}})\right]&=2^{1-s_k}\Tr\left[\rho_k^{0}(\sigma_{{X}}\otimes\sigma_{{X}})\right]\prod_{j=1}^{s_k-1}\left[1+\sqrt{1-\left({\eta^Z_{k,j}}\right)^2}\right].
	\end{aligned}
\end{equation}
Inserting these results into Equation (\ref{IK}) and noting that $\Tr\left[\rho_k^{0}(\sigma_{{Z}}\otimes\sigma_{{Z}})\right]=\Tr\left[\rho_k^{0}(\sigma_{{X}}\otimes\sigma_{{X}})\right]=1$ and $\Tr\left[\rho_k^{0}(\sigma_{{X}}\otimes\sigma_{{Z}})\right]=\Tr\left[\rho_k^{0}(\sigma_{{Z}}\otimes\sigma_{{X}})\right]=0$ we obtain
\begin{equation}
	\begin{aligned}
		&I_s=\frac{1}{2^m}\prod_{k=1}^{m}\left\{\eta_{k,s_k}^Z\cos\theta \Tr\left[\rho_k^{(s_k-1)}(\sigma_{{Z}}\otimes\sigma_{{Z}})\right]+\eta_{k,s_k}^Z\sin\theta \Tr\left[\rho_k^{(s_k-1)}(\sigma_{{Z}}\otimes\sigma_{{X}})\right]\right.\\&\left.+\eta_{k,s_k}^X\cos\theta \Tr\left[\rho_k^{(s_k-1)}(\sigma_{{X}}\otimes\sigma_{{Z}})\right]+\eta_{k,s_k}^X\sin\theta \Tr\left[\rho_k^{(s_k-1)}(\sigma_{{X}}\otimes\sigma_{{X}})\right]\right\}\\	
		&=\frac{1}{2^m}\prod_{k=1}^{m}2^{1-s_k}\left\{\eta_{k,s_k}^Z\cos\theta\prod_{j=1}^{s_k-1}\left[1+\sqrt{1-\left({\eta^X_{k,j}}\right)^2}\right]+\eta_{k,s_k}^X\sin\theta \prod_{j=1}^{s_k-1}\left[1+\sqrt{1-\left({\eta^Z_{k,j}}\right)^2}\right]\right\}.\\	
	\end{aligned}
\end{equation}
Similarly we can write $J_s$ explicitly as,
\begin{equation}
	\begin{aligned}
		&J_s=\frac{1}{2^m}\sum_{y_s}\sum_{a,b_s} (-1)^{a+\sum_k{(b_{k,s_k}+ y_{k,s_k})}} p(a,b_s|1,y_s)\\
		&=\frac{1}{2^m}\prod_{k=1}^{m}\left\{\eta_{k,s_k}^Z\cos\theta \Tr\left[\rho_k^{(s_k-1)}(\sigma_{{Z}}\otimes\sigma_{{Z}})\right]-\eta_{k,s_k}^Z\sin\theta \Tr\left[\rho_k^{(s_k-1)}(\sigma_{{Z}}\otimes\sigma_{{X}})\right]\right.\\&\left.-\eta_{k,s_k}^X\cos\theta \Tr\left[\rho_k^{(s_k-1)}(\sigma_{{X}}\otimes\sigma_{{Z}})\right]+\eta_{k,s_k}^X\sin\theta \Tr\left[\rho_k^{(s_k-1)}(\sigma_{{X}}\otimes\sigma_{{X}})\right]\right\}\\	
		&=\frac{1}{2^m}\prod_{k=1}^{m}2^{1-s_k}\left\{\eta_{k,s_k}^Z\cos\theta\prod_{j=1}^{s_k-1}\left[1+\sqrt{1-\left({\eta^X_{k,j}}\right)^2}\right]+\eta_{k,s_k}^X\sin\theta \prod_{j=1}^{s_k-1}\left[1+\sqrt{1-\left({\eta^Z_{k,j}}\right)^2}\right]\right\}.\\	
		\label{IK2}
	\end{aligned}
\end{equation}

Then we obtain Eq. (4) in the main text, 

\begin{align}\label{eq}
	& S_{s}=\prod_{k=1}^{m}2^{\frac{1-s_k}{m}} \left(\eta^Z_{k,s_k}\cos\theta \prod_{j=1}^{s_k-1}f^X_{k,j}+ \eta^X_{k,s_k}\sin\theta \prod_{j=1}^{s_k-1}f^Z_{k,j}\right)^{\frac{1}{m}}.
\end{align}
where $f^u_{k,j}=1+\sqrt{1-\left(\eta^u_{k,j}\right)^2}$ for $u\in\{Z,X\}$.

\section{Unsharp measurement}

The positive-operator-valued-measures (POVMs) of a qubit system is given as
\begin{equation}
	\begin{aligned}
		\Pi_{0|\vec{r}}=\frac{1}{2}(\openone +\eta \sigma_{\vec{r}}),~~~~\Pi_{1|\vec{r}}=\openone -\Pi_{0|\vec{r}},
		\label{WeakM}
	\end{aligned}
\end{equation}
where $\mathbb{I}$ is the identity operator, $\sigma_{\vec{r}}=\vec{r}\cdot\vec{\sigma}$, $\vec{r}$ is the Bloch vector with $\abs{\vec{r}}=1$, $\vec{\sigma}=(\sigma_X, \sigma_Y, \sigma_Z)$ are Pauli matrices, and $\eta\in[0,1]$. 

\begin{figure*}
	\centering
	\includegraphics[width=0.6\linewidth]{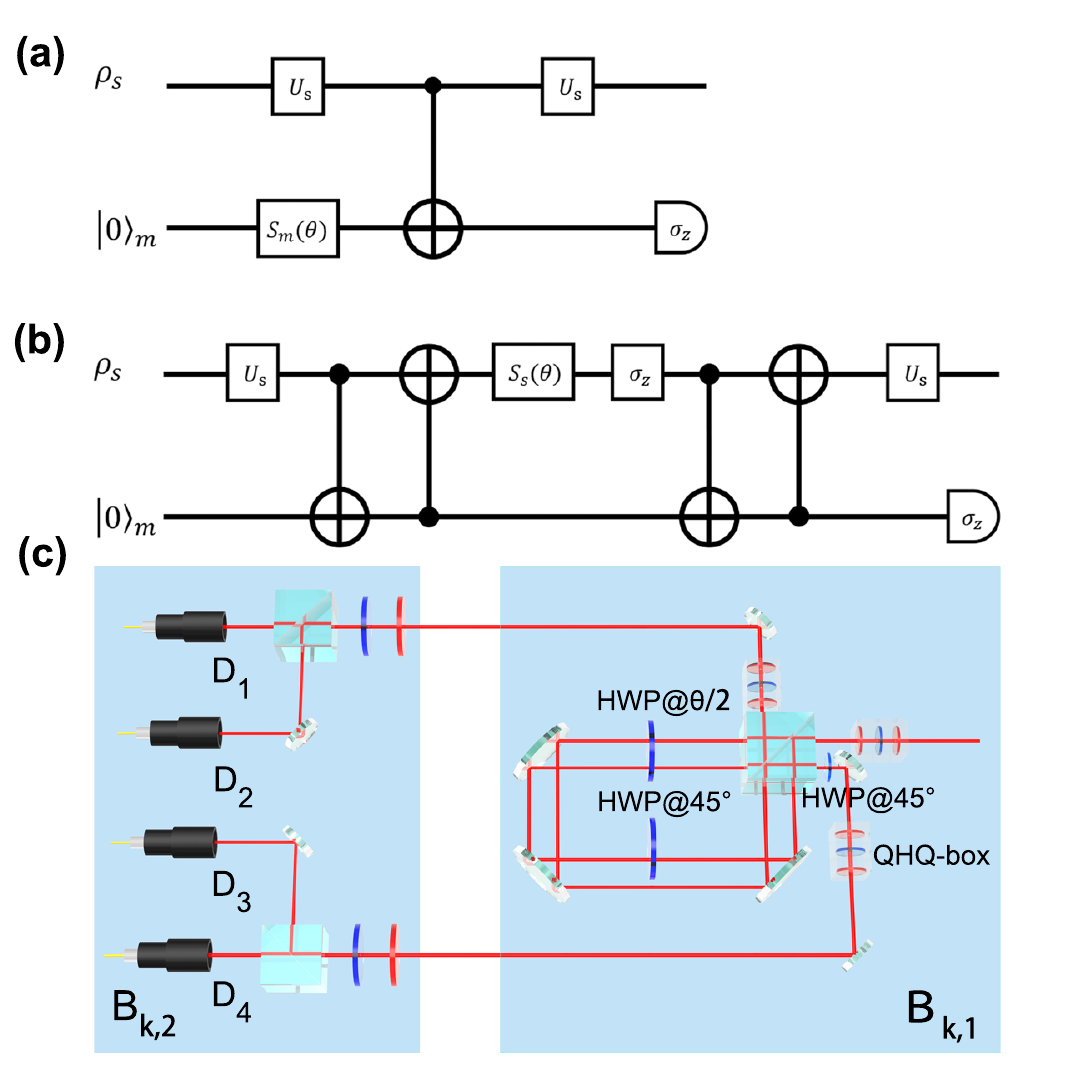}
	\caption{(a) The quantum circuit to realize unsharp measurement Eq. (\ref{WeakM}). 
		(b) The quantum circuit to realize unsharp measurement with a Sagnac interferometer and (c) its optical realization.
	}
	\label{fig:FIGS1}
\end{figure*}

A quantum circuit to implement the unsharp measurement 
is given in Fig. ~\ref{fig:FIGS1}(a), in which the meter qubit $|\Phi_\text{m}\rangle= S_{\text{m}}(\theta) \ket{0_{\text{m}}} = \cos\theta|0_\text{m}\rangle+\sin{\theta}|1_\text{m}\rangle$ couples to the system qubit $\rho_\text{s}$ via a Controlled-Not (C-NOT) gate,
\begin{equation}
	\begin{aligned}
		U^\text{C-NOT}_{\text{sm}}=\ket{0_{\rm s}}\bra{0_{\rm s}}\otimes \mathbb{I}_m+\ket{1_{\rm s}}\bra{1_{\rm s}}\otimes \sigma_{X,\text{m}}.
	\end{aligned}
\end{equation}
The POVMs of the system qubit are given as
\begin{equation}
	\begin{aligned}
		E_{0}&=M^{\dagger}_{0}M_{0}=\frac{1}{2}(\mathbb{I}+\cos 2\theta \sigma_Z)=\Pi_{0|\vec{r}},\\
		E_{1}&=M^{\dagger}_{1}M_{1}=\frac{1}{2}(\mathbb{I}-\cos 2\theta \sigma_Z)=\Pi_{1|\vec{r}},
	\end{aligned}
\end{equation}
with
\begin{equation}
	\begin{aligned}
		M_{0}&=\bra{0_\text{m}}U^\text{C-NOT}_{\text{sm}}(\mathbb{I}_{s}\otimes S_{\text{m}}(\theta))\ket{0_\text{m}}=\cos\theta\ket{0_{\rm s}}\bra{0_{\rm s}}+\sin \theta\ket{1_{\rm s}}\bra{1_{\rm s}},\\
		M_{1}&=\bra{1_\text{m}}U^\text{C-NOT}_{\text{sm}}(\mathbb{I}_{{\rm s}}\otimes S_{\text{m}}(\theta))\ket{0_\text{m}}=\sin\theta\ket{0_{\rm s}}\bra{0_{\rm s}}+\sin \theta\ket{1_{\rm s}}\bra{1_{\rm s}},
	\end{aligned}
	\label{WMO}
\end{equation}
where the unsharp measurement strength is $\eta=\cos 2\theta$. 

Measuring system qubit in other directions $\sigma_{\vec{r}}=\vec{r}\cdot\vec{\sigma}$ are implemented by applying a unitary rotation $U_{\rm s}$ to the system qubit accordingly, for example, $U_{\rm s}$ is Hadamard gate for measurement $\sigma_{X}$, 
\begin{equation}
	\begin{aligned}
		U_{\rm s}=\frac{1}{\sqrt{2}}\begin{pmatrix}
			1 & 1 \\
			1 & -1
		\end{pmatrix},
	\end{aligned}
\end{equation}
and $U_s$ for $\sigma_{Y}$ is 
\begin{equation}
	\begin{aligned}
		U_{\rm s}=\frac{1}{\sqrt{2}}\begin{pmatrix}
			1 & -i \\
			-i & 1
		\end{pmatrix}.
	\end{aligned}
\end{equation}

In this experiment, $B_{k,1}$ encodes system qubit to the polarization state and meter qubit to the path state of photon, with the quantum circuit shown in Fig.~\ref{fig:FIGS1} (b) and experimental realization shown in Fig.~\ref{fig:FIGS1} (c). One can show that the quantum circuit of Fig.~\ref{fig:FIGS1} (b) is equivalent to that of Fig.~\ref{fig:FIGS1} (a). We set meter qubit to state $|0\rangle_\text{m}$ ($|1\rangle_\text{m}$) as photon propagates (counter-) clockwise in the Sagnac interferometer. Below we describe the photon traveling through the Sagnac interferometer with a unitary,

\begin{equation}
	\begin{aligned}
		U^{\rm Sagnac}_{\text{sm}}=  U^{\rm C-NOT}_{\text{ms}}U^{\rm C-NOT}_{\text{sm}} \left[(S_{{\rm s}}(\theta)*\sigma_{Z})\otimes\mathbb{I}_{\text{m}}\right] U^{\rm C-NOT}_{\text{ms}} U^{\rm C-NOT}_{\text{sm}}
	\end{aligned}
\end{equation}

C-NOT gate $U^{\rm C-NOT}_{\text{sm}}$ with meter qubit as target is implemented by the PBS in the Sagnac interferometer. The photon passes the PBS twice and experiences $U^{\rm C-NOT}_{\text{sm}}$ twice. 

C-NOT gate $U^{\rm C-NOT}_{\text{ms}}$ with system qubit as target is implemented by passing photon in state $|0\rangle_\text{m}$ through a HWP oriented at $45^\circ$ (HWP$@45\degree$). The HWP$@45\degree$ inside (outside) the Sagnac interferometer is to implement the first (second) $U^{\rm C-NOT}_{\text{ms}}$.  

Unitary rotation $S(\theta)$ of system qubit is implemented by passing photon through a HWP oriented at $\theta$ (HWP$@\theta/2$) in the Sagnac interferometer. 

$\sigma_{Z}$ is implemented as horizontally polarized photons gain phase $\pi$ upon reflection by a mirror.

We then have  
\begin{equation}
	\begin{aligned}
		U^{\rm Sagnac}_{\text{sm}}=&(\cos \theta\ket{0_{\rm s}}\bra{0_{\rm s}}+\sin \theta\ket{1_{\rm s}}\bra{1_{\rm s}})\otimes \ket{0_\text{m}}\bra{0_\text{m}}
		+(-\sin \theta\ket{0_{\rm s}}\bra{1_{\rm s}}+\cos \theta\ket{1_{\rm s}}\bra{0_{\rm s}})\otimes \ket{0_\text{m}}\bra{1_\text{m}}\\
		+&(\sin \theta\ket{0_{\rm s}}\bra{0_{\rm s}}+\cos \theta\ket{1_{\rm s}}\bra{1_{\rm s}})\otimes \ket{1_\text{m}}\bra{0_\text{m}}
		+(\cos \theta\ket{0_{\rm s}}\bra{1_{\rm s}}-\sin \theta\ket{1_{\rm s}}\bra{0_{\rm s}})\otimes \ket{1_\text{m}}\bra{1_\text{m}}
	\end{aligned}
\end{equation}

\begin{equation}
	\begin{aligned}
		M_{0}&=\bra{0_\text{m}}U^{\rm Sagnac}_{\rm sm}\ket{0_{\rm m}}=\cos\theta\ket{0_{\rm s}}\bra{0_{\rm s}}+\sin \theta\ket{1_{\rm s}}\bra{1_{\rm s}},\\
		M_{1}&=\bra{1_\text{m}}U^{\rm Sagnac}_{\rm sm}\ket{0_\text{m}}=\sin\theta\ket{0_{\rm s}}\bra{0_{\rm s}}+\sin \theta\ket{1_{\rm s}}\bra{1_{\rm s}}.
	\end{aligned}
	\label{WMO_E}
\end{equation}

To implement arbitrary unitary $U_s$, we pass photons sequentially through HWP, QWP, and HWP (QHQ-box), as shown in Fig. \ref{fig:FIGS1}(c).

From Eq.~(\ref{WMO_E}), we can read that photon exiting the Saganc interferometer into the upper or lower path of the setup in Fig. \ref{fig:FIGS1}(c) corresponds to output $b_{k,1}=0$ or $b_{k,1}=1$ of party $B_{k,1}$, respectively. 

Finally, we perform projective polarization measurements by passing photons sequentially through HWP, QWP and PBS, with outcomes $b_{k,2}=0$ if the photon transmits the PBS and $b_{k,2}=1$ if the photon is reflected by the PBS.The single photon detection at detectors $D_1$, $D_2$, $D_3$, or $D_4$ corresponds to one of the four outcome combination $\{b_{k,1},b_{k,2}\}$ as shown in Tab. \ref{tab:Detector}.
\begin{table}[htbp!]
	\caption{ Outcome combination of $\{b_{k,1},b_{k,2}\}$}
	\label{tab:Detector}
	\begin{tabular}{|c|c|c|}
		\hline
		{Detector in Fig. \ref{fig:FIGS1}(c)} & $b_{k,1}$ & $b_{k,2}$  \\
		\hline
		$D_1$ & $0$ & $0$ \\
		\hline
		$D_2$ & $0$ & $1$\\
		\hline
		$D_3$ & $1$ & $1$ \\
		\hline
		$D_4$ & $1$ & $0$ \\
		\hline
	\end{tabular}
\end{table}
\end{document}